\documentclass[a4paper,11pt]{article}

\usepackage{fullpage}
\usepackage{subfigure}
\usepackage{gensymb}
\usepackage{graphicx}

\title{Managing Heterogeneous WSNs in Smart Cities: Challenges and Requirements}

\author{Trang Cao Minh, Boris Bellalta, Simon Oechsner, Ruizhi Liao and Miquel Oliver}

\begin{document}

\maketitle

\begin{abstract}
The dramatic advances in wireless communications and electronics have enabled the development of Wireless Sensor Networks (WSNs). WSNs consist of many affordable and portable sensor nodes for collecting data from the environment. In this article, we address management requirements of WSNs through presenting some key management scenarios in the Smart Cities context, such as intelligent transportation systems, smart grids and smart buildings. The limited resources and heterogeneous characteristics of WSNs pose new challenges in network management, which include the presence of various faults, the difficulty in replacing and repairing a large number of sensor nodes, the existence of an uncertain topology, and the resource allocation. To cope with these challenges, we first discuss advantages and disadvantages of centralized and distributed management approaches and then discuss the benefit of the multilevel management schema. Next, we present in detail the specific features for a WSN management system such as lightweight, self-detection, self-configuration, sharing infrastructure, service monitoring, plug and play, context awareness and interoperability. Finally, we present the required mechanisms for some basic management functions.
\end{abstract}

\textbf{sensor networks, network management, resources allocation, self configuration, context aware, smart cities}

\section{Introduction}

\begin{figure*}[tb]
	\centering
		\includegraphics[width=0.90\textwidth]{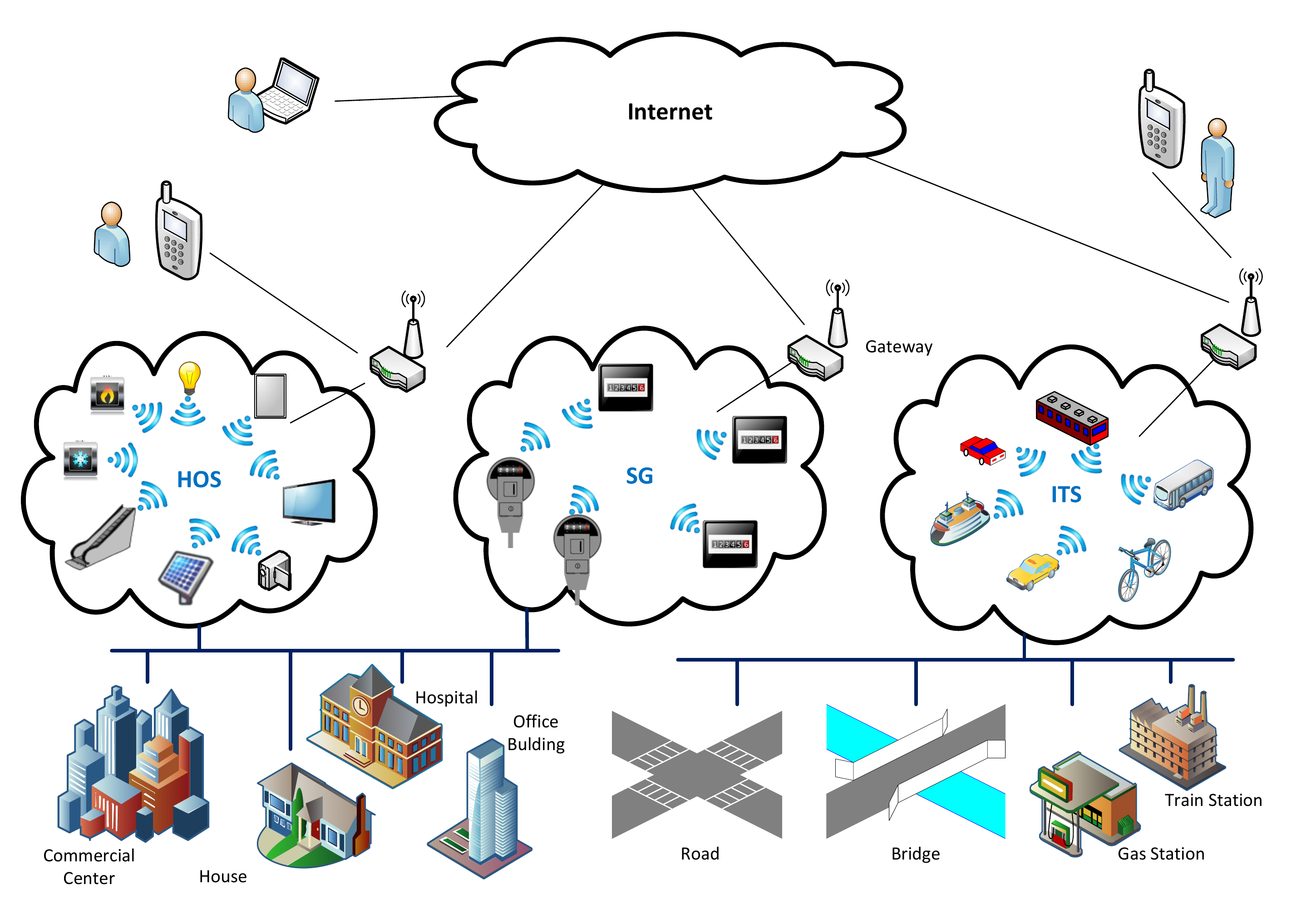}
	\caption{A smart city}
	\label{fig:smartcity}
\end{figure*}

Cities are getting increasingly crowded. Many research groups, in both academia and industry, have recently put huge efforts to make cities smarter to ensure public safety, provide efficient transport, save energy, reduce expenses, and improve quality of life. With advances in wireless communications and MEMS (Micro Electro Mechanical Systems), Smart Cities are becoming a reality. Three most commonly deployed Smart Cites' applications are shown in Figure \ref{fig:smartcity} and can be summarized as follows:

\begin{itemize}
	\item \textbf{Intelligent transportation systems (ITS).} Intelligent transportation systems are applications which apply advances in information and communication technologies with the goal to organize traffic more efficiently, enhance safety and reduce CO2 emissions in transport systems. They can be deployed in vehicles (e.g. car, train, ship, and air plane) and infrastructure (e.g. road, train station, and gas station).
	\item \textbf{Smart Grids (SG).} The growing population has created a greater demand for energy while the limited amount of fossil fuels is diminishing. Additionally, the power grids designed and deployed in the past are not able to cope with current and future needs. To resolve these problems, smarter electrical grids which use information and communication technologies to optimize the energy distribution and to improve the efficiency and productivity of the energy usage are being developed. New smart grids can also help suppliers and consumers to monitor and control the energy usage and costs.
	\item \textbf{Smart Home, Smart Building - Home and Office automation Systems (HOS).} Home and office automation systems interconnect electric devices such as heaters, lights, air conditioners, TVs, computers, alarms, and cameras through a communication network, allowing them to be remotely controlled, monitored or accessed from any room in the building, as well as from any location in the world by Internet. They help people to optimize their living style, arrange the day-to-day schedule, secure a high living quality, and reduce the energy consumption bill.
\end{itemize}

Recently, there are numerous research projects aiming at the development of technologies for such cities, such as Open Cities \cite{opencity} and Smart City Santander \cite{santander}. In the Open Cities project \cite{opencity}, several European cities such as Amsterdam, Barcelona, Berlin, Helsinki, Paris are working on exploring Open and User Driven Innovation methodologies to the Public Sector in a scenario of Future Internet Services for Smart Cities. The Smart City Santander project \cite{santander} focuses on designing, deploying and validating an experimental research facility to support typical applications and services for a smart city. In most of Smart Cities projects, wireless sensor networks play an important role in building instrumented and interconnected urban environments.

Wireless Sensor Networks (WSNs) are made up of small, low power, and low cost automated devices (i.e., sensor nodes), which have the capability of sensing, data processing, and wireless communication. Given these capabilities, WSNs can be deployed in different environments and using a large number of sensor nodes with an affordable cost. Therefore, WSNs are one of the critical components of Smart Cities. For example, in HOS, wireless sensor nodes are used to monitor or detect temperature, light, gas leaks and fire. The output of these sensors can be used to adjust the operation of electric appliances at homes. In ITS, numerous sensors installed on a vehicle can detect obstacles, measure the speed of the leading vehicle, warn impending collisions to the driver and trigger the collision avoidance system when necessary. Infrastructure sensors including induction loops, video and image processing, and microwave radars can be installed on the road to monitor traffic conditions and detect traffic congestion. Several recent studies have examined the feasibility of using WSNs in ITS. Wang et al., in \cite{5676214}, designed and implemented EasiTia, an applicable and cost-effective system for acquiring pervasive traffic information based on WSNs. Recently, Bottero et al. \cite{Bottero201399} have installed and tested a WSN traffic monitoring system in the area of a logistic platform at the Turin's freight village in Italy. In SG, sensors can be embedded in metering devices, placed at both end-points and in the transport network, to monitor and control the energy usage and/or the waste in real time both locally and remotely. They help operators and consumers to manage their energy usage efficiently, reducing their energy bills and optimize delivery networks.

However, one of the biggest limitations of WSNs is usually the scarcity of resources. Sensor nodes are equipped with small batteries (e.g., AA, AAA) with low power capacity. Therefore, sensor nodes are prone to fail due to the battery depletion. These failures can seriously affect the efficiency and the accuracy of the services provided by WSNs. For example, if some sensor nodes on a road are broken, the information about the traffic condition (e.g., number of cars or obstacles) on that road might be wrong. Then, more cars will use that road since they have received inaccurate information about traffic and potential traffic congestion. Therefore, it is essential to have a management system to monitor network operations and node state, to detect and repair faults automatically such that the efficiency and the accuracy of services provided by WSNs are ensured.

In this article, we focus on management problems in WSNs in the Smart Cities context. The rest of the article is organized as follows. We introduce some scenarios to clarify requirements in management for WSNs in Section \ref{sec:scenarios}. A general definition of a management system for WSNs is then presented in the beginning of Section \ref{sec:content}. We discuss management challenges in Section \ref{sec:challenges}, and specific features of a management system for WSNs in Section \ref{sec:feature}. We consider the benefits of the multilevel management architecture for managing WSNs in Section \ref{sec:schema}, and identify requirements of some basic management functions in Section \ref{sec:functions}. Finally, the conclusions are presented in Section \ref{sec:concl}.

\section{Management Scenarios}
\label{sec:scenarios}

In this section, we present some scenarios in different Smart Cities' applications to outline requirements of WSN management systems.

\subsection{Fault or Misbehavior}

There are many faults or misbehaviors which can happen in WSNs. In the following we will discuss two examples.

\textbf{Scenario A:} The electric bill is wrong due to no receiving energy consumption reading caused by errors on node, e.g. battery depletion or sensor broken, or errors on the data delivery path, e.g. network partition or network congestion.

\textbf{Scenario B:} The electric bill is wrong due to receiving incorrect energy consumption readings caused by external attacks, misbehaviors of metering sensor, or errors in network protocols.

These two scenarios provide some different requirements of WSN management systems:

\begin{itemize}
	\item First of all, a WSN management system must be able to determine what has caused the faults. This requires management tasks such as monitoring and fault tracking.
	\item To avoid unexpected effects when a fault occurs, a WSN management system needs to support fault predictability. In other words, it should be able to detect a fault before it occurs by analyzing and validating data including sensing data, network operation logs, etc.
	\item Due to the existence of inevitable faults (e.g. in hardware, in software components, and in network links), a WSN management system needs to detect these faults promptly and reconfigure network operations to ensure the accuracy of the provided service.
\end{itemize}

\subsection{Integration of new sensor nodes or new applications}

During the network's lifetime, there might be the need for deploying new sensor nodes or new applications to replace broken ones, to extend the network, to improve the network performance or to meet new users' requirements. The following are some examples of this situation:

\textbf{Scenario A:} A company wants to deploy a particular security application in its office, which is located in a smart building. This application includes some kinds of sensors such as camera sensors, motion sensors and occupancy sensors to capture any unauthorized activity. There might be also some WSN applications with different types of sensor nodes deployed in the smart building such as the lighting system, the air conditioning system, and the alarm system. Taking advantage of existing resources in the building can reduce the deployment cost of the new user's security application. For example, it can utilize existing occupancy sensors of the lighting systems instead of deploying new ones.

\textbf{Scenario B:} In this scenario, sensor nodes powered by batteries are replaced by ones powered by solar energy in case sensor nodes are located in areas where sunlight is abundant such as green fields or roads. New energy harvesting sensor nodes can eliminate the inconvenience of replacing batteries, and also prolong WSN operational lifetime.

\textbf{Scenario C:} The deployment of a network may include several phases. In each phase, some new sensor nodes may be added to the network.

The management issues that arise in the above scenarios are as follows:

\begin{itemize}
	\item Sensor nodes should be able to support multiple applications which can be owned by multiple users. A WSN management system needs to be able to allocate resources among applications, and also to ensure the privacy of each user.
	\item A WSN management system needs to have a power management mechanism to manage the harvested energy at the harvesting sensor nodes. This mechanism should be able to cooperate with the resource allocation function to align the workload with the energy availability at sensor nodes.
	\item The integration of new sensor nodes or new applications can require a code update process. Due to the large number of nodes in WSNs, a manual update is inefficient. Therefore, a WSN management system should have a remote configuration function.
	\item In order to ensure the compatibility between old sensor nodes and new ones, a WSN management system needs to update the network operations in which new sensor nodes can take part in, such as routing or allocating network resources for applications.
\end{itemize}

\subsection{Quality of Service of WSNs}

Due to the variety of applications in WSNs, the quality of service (QoS) of WSNs varies greatly from application to application. For example, one of the QoS factors is the accuracy. In WSNs which provide information about the physical environment, the accuracy is measured by the discrepancy between the real world value and the provided results. However, in WSNs which are used to decide how to control actuators, the accuracy is measured by the discrepancy between the correct decision and the taken one. Moreover, different QoS factors such as delay and network lifetime may conflict by nature. Two scenarios are introduced to illustrate the conflict among QoS factors.

\textbf{Scenario A:} In the fire detection system in a smart building, important events such as high temperature and the smoke occurrence need to be detected promptly. It requires a high data collecting rate which results in larger energy consumption, more network congestion and higher delays.

\textbf{Scenario B:} There are two WSN applications deployed on a road. The first application is used to detect the traffic congestion. The second one is to detect vehicles that cross a stop line while a red traffic light is on. There is a traffic congestion on the road. To keep live reports, camera sensors need to transmit information of the congestion (e.g. vehicle density, length of congestion, beginning and end of congestion) with a high rate to the sink, which affects the data traffic of the red light application. Information of some cars that violate traffic rules may be lost.

From the above scenarios, a WSN management system must consider the following requirements in order to ensure the required QoS:

\begin{itemize}
	\item The WSN management system should define a QoS model for each application to identify the desired trade-off among QoS factors. It should also identify key QoS factors, if any, that influence the efficiency of the application. For example, the accuracy and the delay are more important for fire detection compared to other factors.
	\item The WSN management system should have a mechanism to monitor the QoS of running services to detect if the QoS of each service is met.
	\item When multiple applications are executed concurrently in a single WSN, the WSN management system should combine the QoS models of all applications and generate a global QoS model to find the general trade-off in case the required QoS of all running applications can not be guaranteed.
\end{itemize}

\subsection{Collaboration among WSNs}

As mentioned above, there are multiple WSNs deployed to support different applications in Smart Cities. However, the WSNs operate independently and belong to different authorities. It would be efficient if the WSNs can cooperate to provide higher services or to improve the network performance. Some scenarios of the collaboration among WSNs are described as below:

\textbf{Scenario A:} A driver wants to find a parking place around a tourist attracting area. The smart parking WSN and the traffic monitoring WSN can collaborate to guide the driver to the most suitable empty parking place without trouble of traffic congestion.

\textbf{Scenario B:} Based on collected information from the traffic monitoring WSN, the pollution monitoring WSN can adjust its data collecting rate correspondingly (e.g. the more traffic the higher rate). This helps the pollution monitoring WSN to keep up-to-date information of pollution while optimizing energy consumption.

To support the collaboration among WSNs, there are new management requirements which a WSN management system needs to take into account.

\begin{itemize}
	\item The WSN management system should be able to analyze and validate data or requests received from other WSNs. Then, it should reallocate network resources to perform received requests.
	\item The WSN management system needs to monitor and evaluate the effects of the collaboration with other WSNs on the network performance. It should be also able to use information of similar previous collaboration requests in handling the current request.
\end{itemize}

\section{A WSN Management System}
\label{sec:content}

Based on the management requirements described previously, a WSN management system can be defined as: \textit{A management system for WSNs is an \textbf{autonomic} framework that keeps the network and the services that the network provides up and running smoothly with as \textbf{little human intervention} as possible, and consumes as \textbf{little resources and energy} as possible. It predicts potential problems, performs operations to avoid or locate them, and self-configures or suggests solutions to solve them. It also allows adjusting network operations and reprogramming nodes remotely. Finally, it supports allocating resources to the services offered by the network.}

\subsection{Challenges in management for WSNs}
\label{sec:challenges}

Due to the scarcity nature of resources and the variety of applications, a WSN management system needs to cope with many challenges which are described in detail as below:

\begin{itemize}
	\item \textbf{Various failures.} In WSNs, faults happen more frequently than other communication networks for many reasons. Firstly, sensor nodes have very limited resources. They are mainly equipped with a small power source (e.g. 2 AA batteries), which only allows them to be continuously active for hundreds hours of operation. In addition, batteries may be defective, hence, shortening node's lifetime. Therefore, sensor nodes are prone to failure due to the depletion of batteries. Secondly, WSNs can be deployed in various environments such as houses, buildings, roads, and rivers. There are a lot of factors which can make sensor nodes or network links fail temporarily or permanently. For examples, nature disasters or traffic accidents can break connections or destroy sensor nodes in one area. Thirdly, WSNs can have a large number of sensor nodes in a small area. In other words, the congestion may occur frequently since multiple nodes may want to transmit packets simultaneously, which leads to packet losses. The characteristics of WSNs make management more challenging. As has been discussed, there are multiple different factors that can cause problems and failures in WSNs. Therefore, figuring out exactly and promptly their causes is extremely challenging for the WSN management system. 
	\item \textbf{Difficult replacement and repair.} WSNs might be deployed in remote, unattended, or hostile environments, which makes difficult, expensive or sometimes impossible to replace or repair broken sensor nodes. In those conditions, potential failures should be identified or predicted before they occur. How sensor nodes are able to predict potential failures and find solutions to prevent them is still an open challenge.
	\item \textbf{Uncertain topology.} Depending on the application, the sensor network topology can be random or pre-determined. For example, in a smart house or a smart building, the location of the sensor nodes is specified. However, in forest fire detection systems, sensor nodes are deployed randomly. Moreover, after the deployment, there may exist a lot of factors that affect the WSN topology, including node faults, different wake up cycles or node movement. For example, node faults might result in broken links and the loss of network connectivity, or sensor nodes can wake up at different periods due to mis-configured or faulty network protocols. In some applications, the sink or sensor nodes are placed on movable objects such as a patient, a vehicle or an animal, resulting in a changing network topology. When the network topology is uncertain, keeping up-to-date network information is more costly since the WSN management system has to monitor frequently the network state. Moreover, management data could be also lost due to a change of the management data forwarding paths. It would result in the degradation of the efficiency of the WSN management system.
	\item \textbf{Resource allocation among heterogeneous sensor nodes.} Traditional WSNs are designed to support a single application that belongs to a single user. However, with the rapid development of MEMS technology, there are more differentiated types of sensor devices with different energy capacity and functionality. This results in the emergence of heterogeneous WSNs that consist of several different types of sensor nodes and, in addition, each sensor node may support multiple applications. For example, in a smart business building, the owner may deploy a WSN which supports multiple applications, including temperature and humidity monitoring, structure health monitoring and security alarms. Using a single network to interconnect all nodes can reduce the deployment and maintenance costs since each node can run several applications. However, that situation also raises new challenges for the network management, such as how to allocate the network resources to different applications, how nodes collect and transmit measured data from different nodes and applications to the sink efficiently, and how to keep the energy consumption as low as possible. Furthermore, recent advances in technologies enable sensor nodes to collect and use energy from the environment \cite{5522465}, e.g. light, differences in temperature, or linear motion instead of batteries. However, the availability of the harvested energy varies with time in a non deterministic manner. For example, the energy extracted from a solar panel depends on the maximum solar radiation and varies during a day. Moreover, different nodes may have different harvesting opportunities. For instance, sensor nodes placed at abundant sunlight areas can gather more energy than ones in shaded areas. Therefore, it is difficult to allocate tasks to the harvesting nodes since they do not have a stable energy source.
\end{itemize}

% ------------------------------------------------------
% ------------------------------------------------------
% ------------------------------------------------------
% ------------------------------------------------------

\subsection{Specific Features}
\label{sec:feature}

In this section, we discuss some specific features of a management system for WSNs to cope with the challenges described above.

\subsubsection{Lightweight}

Since sensor nodes have limited resources, a management system for WSNs should be as lightweight as possible. The management functions and the management process should only occupy a small memory size. There should be a trade-off between the network traffic generated by a management process and the benefit derived from it.

\subsubsection{Self-detection}

There are a variety of faults in WSNs. Simple faults which are caused by hardware error or battery depletion should be detected locally by every sensor node. A couple of simple faults from different nodes can lead to a complex fault (e.g. network congestion or network partition). Complex faults can have a high probability to cause a degradation on the network performance. Therefore, sensor nodes should be able to collaborate to detect complex faults from simple faults.

\subsubsection{Self-configuration}

Operation of sensor nodes should be optimized and able to adapt autonomously to the changes in resources and application requirements to prolong the network lifetime and prevent possible faults. For example, sensor nodes in the same sensing area can collect and transmit sensed data alternately. When a sensor node detects a fault, it should notify other nodes. Depending on the importance of the fault, sensor nodes should be able to adjust their operation to reduce negative effects caused by that fault. For example, if the sensing component of a sensor node is broken, it can have a more important role in the forwarding path since it does not need to collect data from the environment. Therefore, other nodes can change their routing table to use that node as the forwarding node.

\subsubsection{Sharing infrastructure}

The deployment and maintenance of large WSNs with thousands of nodes require a high cost and huge effort. In case many WSNs are deployed in the same area, it would be efficient if they share their resources to support multiple applications from the different authorities. This is clearly seen in the two following examples:

\begin{itemize}
\item \textbf{Single application}. In a smart building, both the lighting system and the security system use occupancy sensors in rooms and corridors. In the lighting system, occupancy sensors are used to turn on/off the light depending on the presence of persons in the room. In the security system, they are used to start monitoring. Much of the same area is covered by both systems.
\item \textbf{Single authority}. In a smart city, there are several organizations (e.g. police, highway agency, and local city authorities) that need to deploy their own camera networks on the roads. However, these networks can cover the same areas and therefore, they may generate redundant information.
\end{itemize}

Therefore, it would be beneficial to have a single infrastructure supporting multiple applications. The shared infrastructure can include many different types of sensor nodes, in which some nodes support multiple applications. The management system of such an infrastructure should be able to allocate resources among applications to optimize the network performance. As the infrastructure can be shared by multiple authorities, it needs an access classification mechanism that assigns different privileges to different authorities to ensure the privacy.

\subsubsection{Service Monitoring}

The QoS of running services in WSNs should be monitored and evaluated periodically to detect whether it meets the predefined requirements. A management system for WSNs should also provide detailed information about the availability of the running services.

\subsubsection{Context aware}

As mentioned above, a WSN should support multiple applications. Since each application has different requirements in terms of network resources, and both the application requirements and the network resources change over time, the network behavior should adapt to optimize its performance. During the network lifetime, there might be some situations that can be predicted before they happen. For example, there are more customers at commercial centers during weekends than weekdays. Then, to offer customers a comfortable shopping environment, the commercial centers' smart systems (e.g. lighting and air conditioning) may increase the operating power and the operating frequency autonomously every weekend. In such cases, a management system for WSNs should be able to use information of the handled changes to process the current ones if they are similar. For example, the adjustment of the operating power and the operating frequency of the smart systems last weekend can be applied to the current weekend if the number of guests and the outside air temperature are similar.

\subsubsection{Plug and Play}

Due to the heterogeneity of WSNs, management functions should be as independent as possible from hardware, network protocols and user applications. The same management function should be able to work with different applications, different operating systems, different hardware and different network protocols. This would help to reduce the developing cost. In order to achieve this feature, management functions should be parameterizable and configurable in order to be able to integrate easily with other network protocols, hardware functions and different applications. In order to optimize the memory usage, management functions should be only added to a node when they are needed, and therefore, they should be able to be added or removed easily.

\subsubsection{Interoperability}

In order to enable the collaboration among different WSNs, a management system for WSNs should support interoperability. It means the WSN management system is able to communicate and exchange data with ones of other WSNs. Data standards and public interfaces should be defined and unified among the authorities of WSNs to facilitate the collaboration.

\subsection{Multilevel Management Schema}
\label{sec:schema}

One of the most important aspects of the design of a network management system is its architecture. Since the size of WSNs can range from a small number to thousands of sensor nodes, a management system for WSNs should be scalable. It should work in both small and large networks. Additionally, it should support adding or removing nodes, protocols and applications easily without affecting the on-going network operation and the perceived network performance.

\subsubsection*{Centralized approach}

A centralized management server that processes the management data and take management decisions may be the best option for small networks. This central management server collects information from all sensor nodes and controls the entire WSN operation \cite{Ramanathan:2005:SSN:1098918.1098946,Zhao:2009:HWH:1683300.1683804}. Due to its abundant resources and the global knowledge of the WSN, it can perform complex management tasks and provide accurate management decisions. Complex management tasks are actions that require a high amount of resources and global knowledge of the network. For instance, controlling the network topology is a complex task. However, for large WSNs, it is difficult and costly to keep the management data from all the nodes in the network up to date. Firstly, sensor nodes cannot send management data frequently to the central server due to the high communication overheads of the multi-hop forwarding. Secondly, transmitting management data frequently to the central server increases the traffic load of the nodes close to the sink, which can cause network congestion and lead to high packet losses.

\subsubsection*{Distributed approach}

Distributed management approaches are more suitable than centralized ones in large scale networks. Management decisions are taken by multiple responsible management nodes \cite{Ruiz2003,Cha2007}. Each responsible management node controls part of the network (i.e., a group of nodes), and can cooperate with other responsible management nodes if needed. However, the main disadvantage of distributed approaches is that responsible management nodes do not have a global view of the network, as they only know the state of their respective subnetwork. Therefore, although their management decisions can be effective for their local subnetwork, they can affect negatively the operation of the overall network. For example, some nodes are turned off by a responsible management node in its subnetwork to optimize the resource usage. If those nodes happen to be the only connections to the rest of the network, the whole network will be severely affected. The cooperation among responsible management nodes can improve somewhat this issue, but it may imply a high overhead. For example, responsible management nodes can cooperate to decide which nodes are going to sleep without affecting the network connectivity. However, if the WSN has a lot of subnetworks, the number of management packets exchanged among responsible management nodes will be high. Furthermore, not all responsible management nodes have rich power sources or strong processing capabilities, which means that the number of management functions and the complexity of management functions at those stations are limited.

\subsubsection*{Hybrid solution}

\begin{figure*}[tb]
	\centering
		\includegraphics[width=0.80\textwidth]{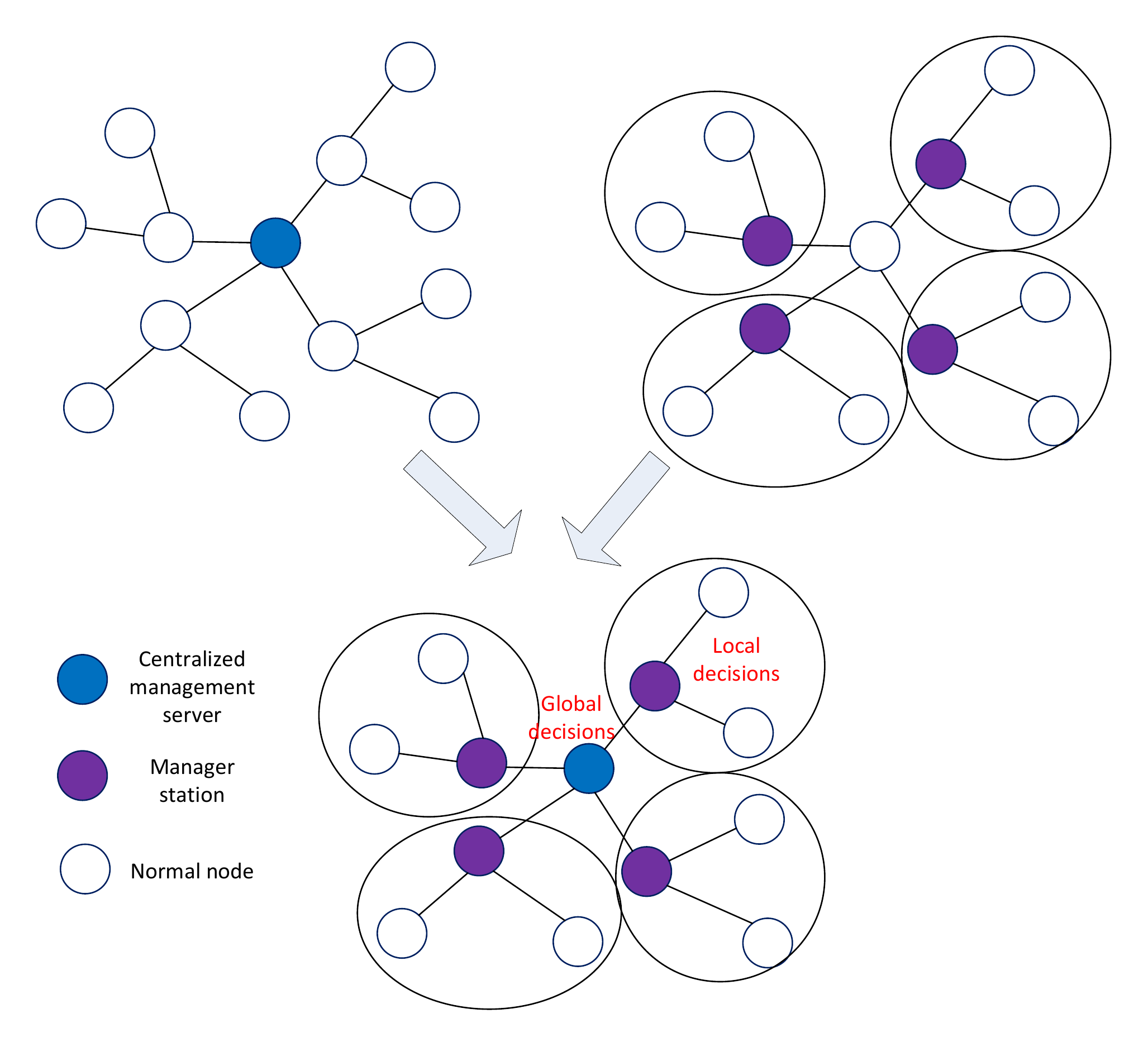}
	\caption{A hybrid management architecture for WSN management systems}
	\label{fig:architecture}
\end{figure*}

Both centralized and distributed approaches have advantages and disadvantages. To take advantages of both approaches, a hybrid management architecture could be designed for WSNs (Fig. \ref{fig:architecture}). In a simple term, a hybrid management architecture consists of both centralized management server and responsible management nodes to perform management functions based on the complexity and the cost required by these functions. For example, a recent work \cite{Minh2012} proposed a multilevel management system for WSNs, in which every node, depending on its resources, participates in the management process at different levels. The approach allows some special nodes to manage a group of adjacent nodes. In small networks, these special nodes can be the sink nodes, and make the management approach centralized. In case of large networks, these special nodes can be selected from sensor nodes which have abundant resources. They can perform simple management tasks locally to reduce management traffic to the sink. Complex management tasks would be performed by the sink, or an external server who has high processing capabilities.

These are the following advantages of a hybrid management architecture:

\begin{itemize}
\item \textbf{Reliable.} It can detect, handle, and isolate faults locally without affecting the functioning of the rest of the network. In can also provide accurate management decisions due to the existence of the centralized management server.
\item \textbf{Scalable.} It is easy to increase the size of the network by adding new sensor nodes, without affecting the current network operation.
\item \textbf{Flexible.} According to the changes in application requirements, the network topology, and the network resources, sensor nodes can have different management roles. For example, when the remaining battery of a responsible management node is low, one of its neighbors can become a new responsible management node to ensure the execution of management tasks in that area.
\item \textbf{Effective.} Responsible management nodes can be selected based on their resources or their roles in network (e.g. cluster head or parent node in routing tree). Therefore, the delay of handling management decisions and management overhead can be minimized.
\end{itemize}

Although the hybrid solution have many advantages, the design of a hybrid management architecture is complex. It requires to have an efficient algorithm to choose the responsible management nodes. However, there are a lot of clustering algorithms developed for WSNs \cite{Abbasi20072826} which can be used to select responsible management nodes. Another disadvantage of the hybrid solution is the management overhead. Exchanging management data can cause high traffic and energy consumption. Due to the lack of resources in WSNs, a hybrid management architecture should be designed to ensure the trade-off between the management overhead and the efficiency of the management system. The number of exchanging management data should be minimized while it still ensures the accuracy of management decisions.

\subsection{Desired Management Functions}
\label{sec:functions}

\subsubsection{Monitoring}

Monitoring is one of the most important management functions. It is responsible for collecting the information required by the management system to monitor the running status of the network, including the network topology, the remaining energy of nodes in the network, and the quality of provided services, among others.

\begin{figure*}%[thp]
\centering{
    \subfigure[]{ \label{fig:monitor1.pdf}
        \includegraphics[width=.3\textwidth]{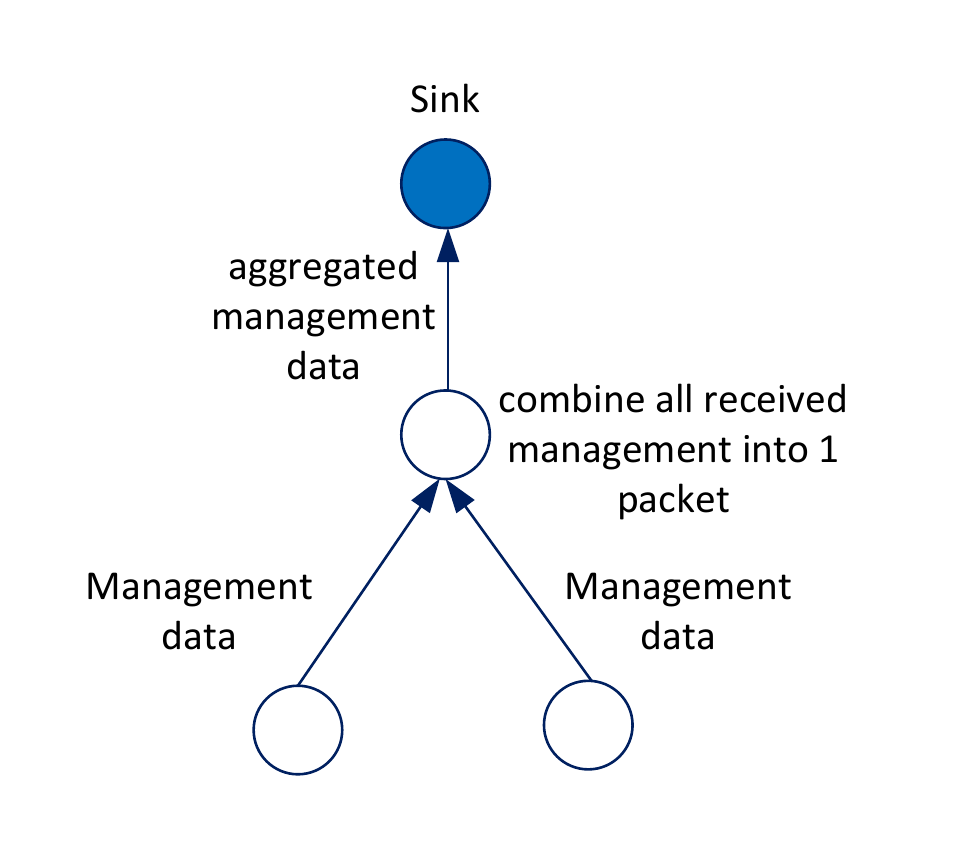}
    }
    \subfigure[]{ \label{fig:monitor2.pdf}
        \includegraphics[width=.4\textwidth]{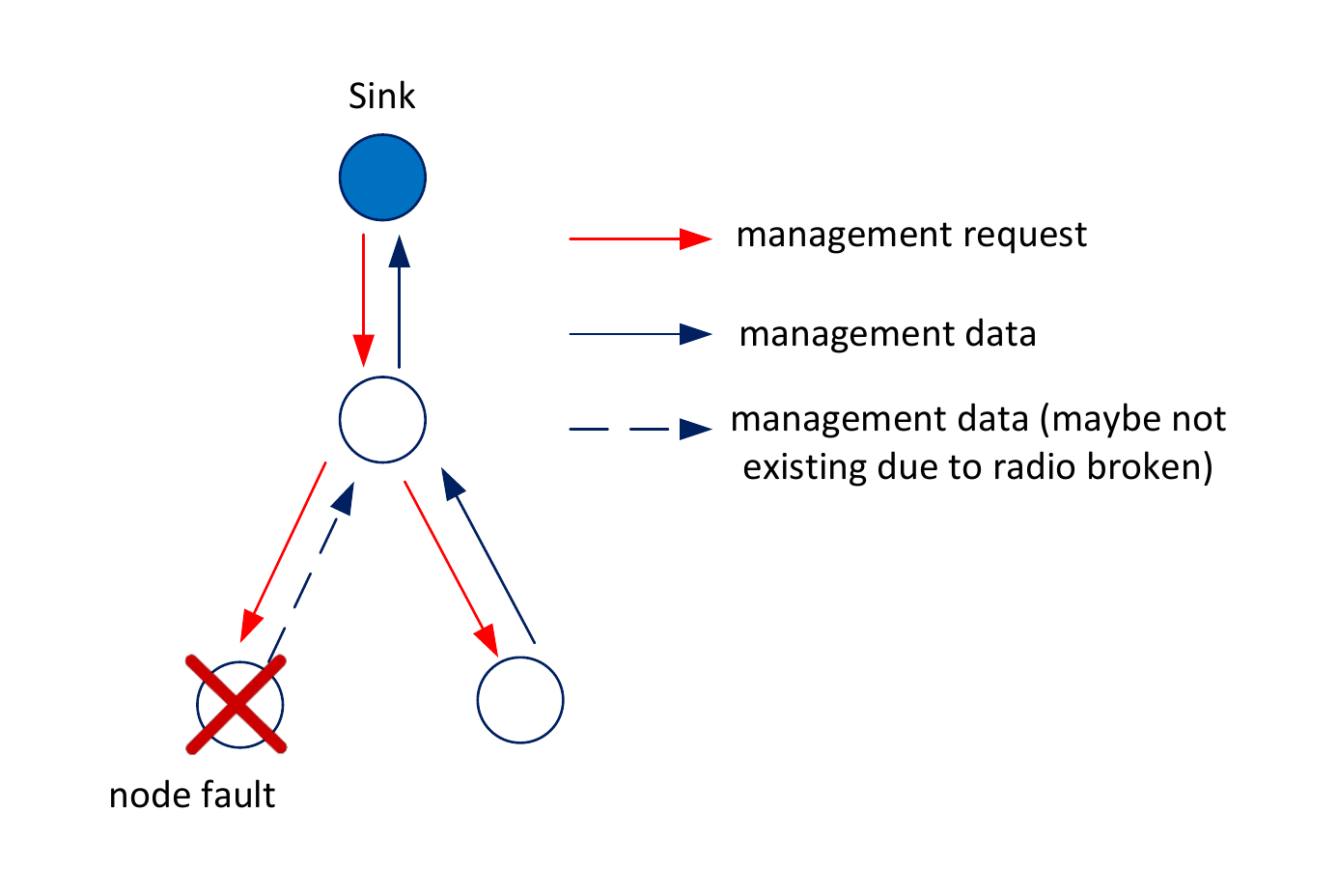}
    }
    \subfigure[]{ \label{fig:monitor3.pdf}
        \includegraphics[width=.2\textwidth]{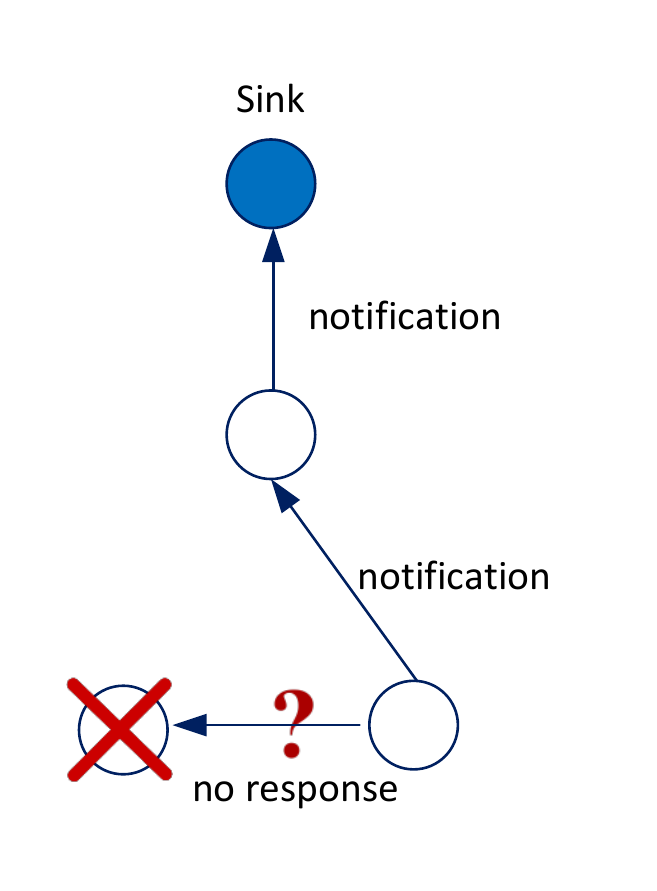}
    }
 }
\caption{Monitoring mechanisms: (a) Periodic, (b) On demand, (c) Event based monitoring}
\label{fig:monitoring}
\end{figure*}

There are three basic monitoring mechanisms to be considered:

\begin{itemize}
	\item \textbf{Periodic.} Sensor nodes should transmit information about its resources, e.g. the remaining battery level, the operation logs of network protocols, or the number of active applications to the responsible management nodes or to the sink periodically. This would help to predict and diagnose possible problems and potential failures. Management information should be aggregated when it is transmitted to reduce the management overhead. An example of the aggregation of management information is shown in Figure \ref{fig:monitor1.pdf}. In addition, the transmission of management information has to face with a dilemma. While the frequent transmission causes a lot of energy consumption, the opposite results in late fault detection. Therefore, the period of time between two transmitted reports by each node must be optimized to provide the required information on time without overloading the network.
	\item \textbf{On demand.} A management system for WSNs should allow to collect management data when it is needed. The Figure \ref{fig:monitor2.pdf} illustrates a simple example of the on-demand monitoring mechanism. The sink detects a fault (e.g. the packet delivery rate suddenly drops below a threshold), hence, it broadcasts management requests to the network to collect essential information to figure out what happened and why this issue has occurred.
	\item \textbf{Event based monitoring.} A sensor node should be able to send a notice to its responsible management node or to the sink as soon as it detects a fault. It helps responsible management nodes and the sink to react promptly in case the detected fault is serious. As illustrated in Figure \ref{fig:monitor3.pdf}, a sensor node detects that one of its neighbors may be broken because it has not received any information from this neighbor for a while. Once the sensor node notified its responsible management node of that unexpected event, it will investigate and evaluate the importance of that suspect node. If the suspect node has an important role in the communication paths or in some applications, the responsible management node will have to reconfigure the routes or reallocate resources to prevent a degradation in the network performance.
\end{itemize}

These monitoring mechanisms should work concurrently in the WSN management system to ensure there is no missing and unsolved problems. They should be able to cooperate in some management processes. For example, the WSN management system detects a fault based on the collected information from the periodic monitoring mechanism. It can trigger the on demand monitoring mechanism to investigate what happened. 

Some examples of monitoring approaches are proposed in \cite{5462033,WinnieLouisLee2006}. In \cite{5462033}, Liu et. al. proposed a two tier structure where nodes in the lower tier send status reports to nodes in the higher one. Each node at the higher tier makes local decisions based on the received data, and forwards its decisions towards the sink. Lee et al. in \cite{WinnieLouisLee2006} presents a schedule-driven MAC protocol to collect and disseminate management data, to and from sensor nodes in a data gathering tree.

\subsubsection{Resource Allocation}

An efficient resource allocation schema is necessary when multiple tasks, from different applications, run simultaneously in the same node and the same network. The resource allocation schema is responsible for assigning network resources to different applications in order to ensure the quality of provided services while prolonging the network lifetime. The first process of the resource allocation schema is validation, which includes the following mechanisms:

\begin{itemize}
	\item \textbf{Access Verification.} This mechanism verifies that the users who request the task are authorized users.
	\item \textbf{Ability Validation.} This mechanism checks if the network, a sensor node or a group of nodes can satisfy the QoS required by the new task. It also checks if the new task affects the QoS of the existing tasks.
\end{itemize}

\begin{figure*}[tb]
	\centering
		\includegraphics[width=0.90\textwidth]{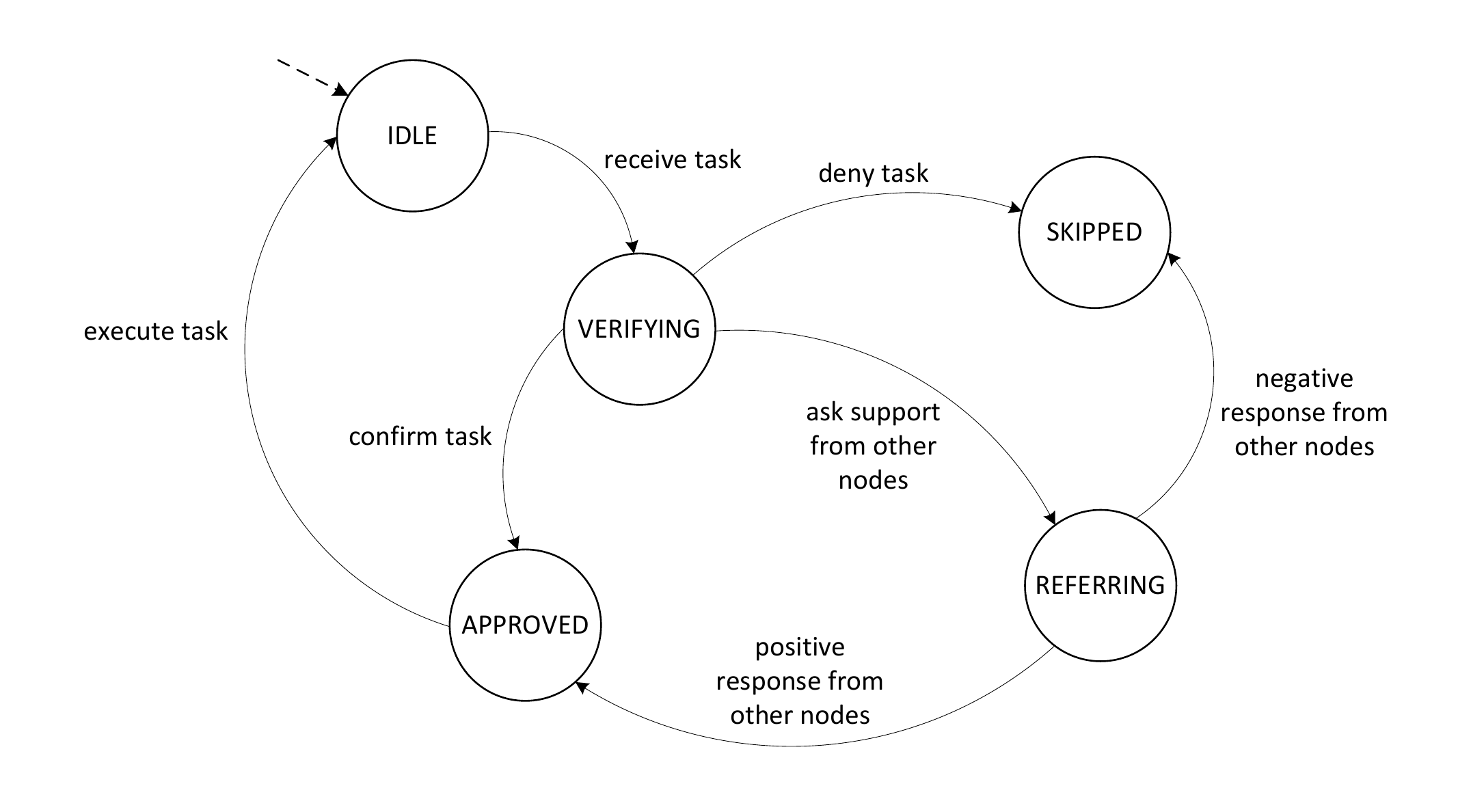}
	\caption{Validation process}
	\label{fig:resalloc}
\end{figure*}

A simple illustration of the validation process is shown in Figure \ref{fig:resalloc}. When a node receives a new task, it checks its resources to verify if it can execute that task, has to deny the request, or needs to ask for supports from other nodes.

In case the new task is approved, it becomes an active task. Before executing the new task, the WSN management system should be able to combine the requirements of all active tasks. For example, consider that there is a task that collects temperature measurements if they are in the $[20, 30]\degree C$ range. Then, a new arriving task requests to collect temperature measurements if they are in the $[15, 25]\degree C$ range. In such case, the new task should report the temperature if it is in the $[15, 30]\degree C$ range. We refer to this new task as an aggregated task, as it aggregates in a single task the existing and the new one.

Before a new task is implemented, it is necessary to have a scheduling mechanism to allocate active tasks in a way that sensor nodes are able to collect and disseminate its outputs intelligently to minimize the number of generated packets, and hence achieve bandwidth and energy efficiency. In \cite{Minh2012}, Minh et al. proposed a resource allocating mechanism where each node checks whether it can perform the requested operation based on its resources. Manager nodes coordinate tasks in groups of adjacent nodes based on the node's resources to avoid that multiple nodes execute the same task, hence, reducing the data duplication and the energy consumption.

Similarly, when a node is added or removed from the WSN, the WSNs management system should be able to re-allocate the network resources to ensure that they are used efficiently. In case the network performance is affected when those changes are applied  (e.g. the connectivity of network are lost when the removed nodes have important role in the routing path), the WSNs management system should be able to suggest feasible solutions to resolve the problem.

\subsubsection{Fault Management}

The Fault Management system is responsible for the analysis of the causes and the search of solutions when a fault occurs. When the Monitoring function detects a new fault, the WSN management system should perform the following steps:

\begin{itemize}
	\item \textbf{Fault Evaluation.} This step carries out an initial assessment on the importance of the fault. It predicts the impact of the detected fault on the system operation and performance. If the fault is serious, it will continue investigating. Otherwise, it might record the fault to process it later, or simply ignore it. For example, if the sensing component of a sensor node is broken, this does not affect the data collection if there are other sensor nodes in the same area. However, these other nodes might have the same fault in the future. Therefore, fault records can be used to identify the number of working nodes and send warnings when needed.
	\item \textbf{Fault Tracking.} This function is used to collect more information from the network to determine why the fault happened.
	\item \textbf{Fault Solution.} This function issues a solution to fix the fault. It should include a mechanism to predict the effect of the solution in the network operation and its performance.
\end{itemize}

One of the important tools for the fault management is the debugging tool. It is necessary to support both pre- and post-deployment debugging tools \cite{Ramanathan:2005:SSN:1098918.1098946,5368155}. The pre-deployment debugging tool allows for the prediction of possible failures, and also has the ability to handle them by simulating WSN operations. The post-deployment debugging tool is used to locate failures during the run time.

Since sensor nodes have very limited resources, it is costly to collect management data from all nodes in multi-hop networks. Therefore, there should be a simulation environment that reproduces the real network deployment (e.g. simulating network operations with the same number of nodes in a similar topology), which can be used to reproduce failures and evaluate the impact in a quick and cheap way.

\subsubsection{Configuration}

This function is used to reconfigure node's operation and update new code. There should be three levels of the node's configuration:

\begin{itemize}
	\item Self-Configuring. Before any simple fault occurs, a sensor node should be able to self-configure its operation to avoid it. For example, harvesting nodes change their duty cycle according to the energy gathering rate to balance between the harvested energy and the consumed.
	\item Cooperative Configuring. Sensor nodes should be able to cooperate to configure their operation to avoid potential network failures. For example, in some cases, a node that has a low battery level can have an important role on a forwarding path. If that node is broken, it can cause a network partition. However, if it changes its duty cycle or if it decreases its transmission power, the lifetime of the node can be prolonged but the routing path can be affected negatively. Therefore, neighbor nodes can decide to change their routing tables to limit the amount of network traffic that goes through that node. Additionally, they can cooperate to find the subset of nodes which are active in routing or in providing a specific service at a given time, while the others are inactive to reduce energy consumption \cite{Minh2012,1318596}.
	\item Remote Configuring. The sink should be able to reconfigure the whole network when necessary. For example, it can change the duty cycle of all nodes, adjust the operating parameters of the network protocols to improve the network performance, and deploy new code to update or change the network operation \cite{Hagedorn:2008:RDO:1371607.1372755}.
\end{itemize}

\section{Conclusion}
\label{sec:concl}

The heterogeneity of technologies and applications, as well as the specific requirements and limitations of WSNs, make necessary to deploy a smart management system to guarantee the correct operation and performance of the sensor networks. In this article, we have introduced a set of relevant management scenarios for WSNs, in the context of Smart Cities. Through those scenarios we have justified the requirements of a management system for WSNs. We have then presented the challenges that a management system for WSNs needs to take into account. We have also discussed the advantages and the disadvantages of both centralized and distributed management approaches, showing that an hybrid management architecture might be the ideal solution to achieve the required functionalities in a scalable way.

In particular, we have introduced a general definition of a management system for WSNs which considers their specific characteristics, such as their limited processor, memory and battery resources. Specific features of a management system for WSNs are also discussed in detail. Finally, we have presented the requirements of a set of key mechanisms in a management system for WSNs, such as the monitoring mechanism, the resource allocation, the fault management and the node's configuration management.

\end{document}